\documentclass{appolb}
\usepackage{graphicx}

\begin{document}
\title{Spin Fluctuations and Superconductivity around the Magnetic Instability%
}


\author{T\^{o}ru Moriya
\address{Department of Physics, Faculty of Science and Technology \\
Tokyo University of Science, Noda, 278-8510}
}
\maketitle


\begin{abstract}
We summarize the present status of the theories of spin fluctuations in dealing with 
the anomalous or non-Fermi liquid behavior and unconventional superconductivity 
in strongly correlated electron systems around their magnetic instabilities or quantum critical points. 
Arguments are given to indicate that the spin fluctuation mechanisms is the common origin of superconductivity 
in heavy electron systems, 2-dimensional organic conductors and high-$T_{\rm c}$ cuprates.  
\end{abstract}

\PACS{74.20.Mn, 74.25.-q, 75.40.Gb}

 
\section{Introduction}

\begin{figure}[!ht]
\begin{center}
\includegraphics[width=1.\textwidth]{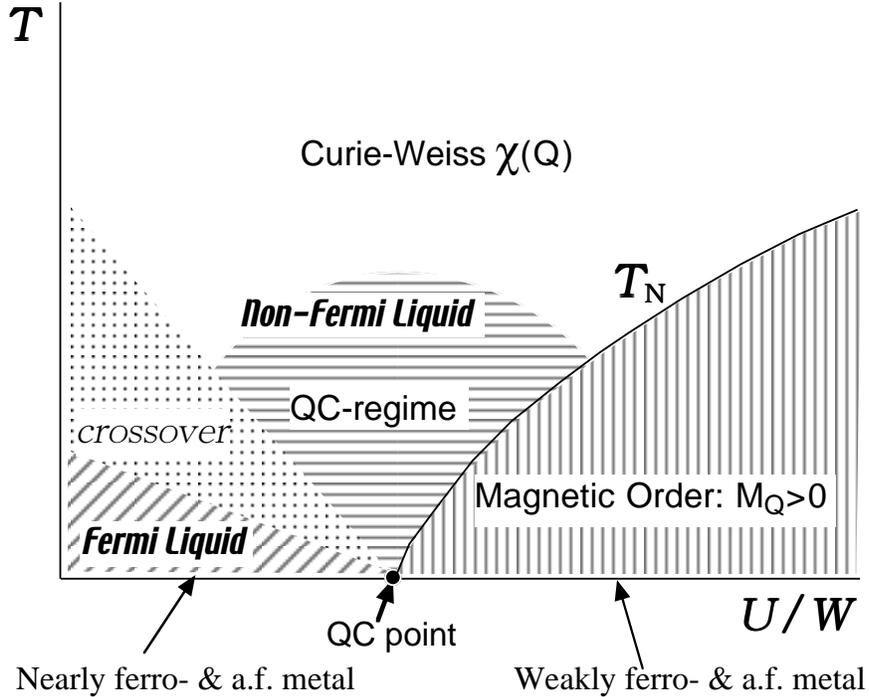}
\vspace*{-2.5cm}
\end{center}
\caption{A schematic phase diagram around the magnetic quantum critical (QC) point.}
\label{fig1}
\end{figure}

Anomalous physical properties and unconventional superconductivity around the magnetic instability 
have been the subjects of intensive current interests. Let us first look at Fig. ~\ref{fig1}, 
showing a phase diagram around the magnetic instability or quantum critical (QC) point which separates 
the paramagnetic Fermi liquid(FL) phase and the magnetic ordered phase with the ordering vector $Q$ at $T = 0$. 

Substances belonging to this area have been called nearly and weakly ferro- and antiferromagnetic metals. 
It seems that the magnetism in this area was called first attention by the discovery of weak itinerant ferromagnets such 
as ZrZn$_2$ and Sc$_3$In~\cite{1}. 
The Curie-Weiss(CW) magnetic susceptibility observed in these systems posed a serious challenge to theorists, 
urging them to consider a new mechanism without local moments. 
For nearly ferromagnetic metals, on the other hand, attention was attracted by a prediction based on the paramagnon 
theory that the effective mass in the FL regime diverges logarithmically as one approaches the QC point~\cite{2,3}.  

Since then the theory of spin fluctuations has been developed to deal with the problems in the entire area 
of the phase diagram shown in Fig. ~\ref{fig1}. 
The self-consistent quantum mechanical theory of the coupled modes of spin fluctuations or 
the self-consistent renormalization (SCR) theory not only succeeded in explaining the 
Curie-Weiss susceptibility of $\chi_{Q}$ but also predicted various singular behaviors of 
the physical quantities in the QC regime, which have been confirmed by subsequent experimental 
investigations~\cite{4,5}.

The discovery of superconductivity of the materials in this area, heavy electron systems~\cite{6}, 
organic conductors~\cite{7}
and in particular high-$T_{\rm c}$ cuprates ~\cite{8}
has renewed and enormously enhanced the interests in this area.  

Upon discovery of the high temperature superconductivity of the cuprates the attention of investigators was 
first concentrated on the anomalous normal state properties or the so-called non-Fermi liquid properties 
of these substances where strongly correlated quasi 2-dimensional(2D) electrons in each CuO$_2$ 
layer were considered to play a predominant role in conduction~\cite{9}.

The spin fluctuation theory was extended to cover 2D systems and it was found that the results in the QC and CW 
regimes were consistent with the observed non-FL behaviors of the cuprates~\cite{10}. 
The experimental results were successfully analyzed in terms of the parametrized SCR theory 
and the parameter values were estimated. 

In the next step the spin fluctuation-induced superconductivity was studied by using the spin fluctuations 
thus determined quantitatively. Using both weak coupling ~\cite{10,11}
and then strong coupling theories~\cite{12,13}
the superconductivity of $d$-$x^2-y^2$ symmetry was derived with the values of the transition temperature $T_{\rm c}$ 
consistent with the experimental results. The $d$-wave symmetry of the superconducting order parameter of 
the cuprates was confirmed by experimental investigations in the following years~\cite{14}.

Subsequent studies on the relation between $T_{\rm c}$ 
and the spin fluctuation parameters showed that the value of $T_{\rm c}$
was roughly proportional to the parameter $T_0$ indicating the energy spread of the spin fluctuations ~\cite{15,16} 
This relation may be extended to certain 3D systems and it was shown that the plots of experimental values of 
$T_{\rm c}$ against $T_0$ for cuprates and heavy electron systems came around a straight line~\cite{17}. 
This result suggests the common origin of superconductivity in these systems.   

Theoretical studies using the Hubbard and the $d$-$p$ models have been advanced not only for the 
cuprates~\cite{18,19,20,21,22,23,24,25}
but also for the 2D organic superconductors~\cite{26,27,28}
by using the fluctuation exchange (FLEX) approximation which may be regarded as the simplest self-consistent theory 
for the coupled modes of spin fluctuations.  
The results turned out to be successful indicating that the Hubbard model with proper transfer matrices and a proper 
electron occupation is a good model for these systems. 

A phase diagram is then calculated within the same approximation in a parameter space of 
the Hubbard model containing the parameter values for both the cuprates and the organic superconductors. 
It was found that both of them belong to a continuous region of a superconducting phase, 
indicating their common physical origin~\cite{29}.

The theoretical approaches discussed so far are the one around the magnetic instability or QC point, 
which belongs to the intermediate coupling regime. 
As will be discussed later in this article there have been growing pieces of evidence to indicate that the high-$T_{\rm c}$ 
cuprates are in the intermediate coupling regime and the spin fluctuation theory is reasonably applied to the main part 
of the problem. 

At this point we would like to mention the pseudo-gap phenomena observed in the under-hole-doped cuprates as 
a still remaining outstanding subject in the high-$T_{\rm c}$ problems.  
Although the physics of doped Mott insulators in the entire concentration range, including the pseudogap phenomena, 
seems still to be worked out theoretically, it is fortunate that at least a substantial part of 
the  SC phase of the cuprates is around the QC regime and is described essentially in terms of the theories 
of spin fluctuations.      

In the following sections we summarize very briefly the developments 
in the theory of spin fluctuations as applied to magnetism and superconductivity and discuss to 
what extent we now understand these problems.   

\section{Developments of the spin fluctuation theory \\
		      {\it Quantum critical and non-Fermi liquid behaviors}}

The dynamical susceptibility around the magnetic instability for the state with the ordering vector $Q$ 
may be expressed for small $q$ and $\omega/q^{\theta}$ as follows:
\begin{eqnarray}
     \frac{1}{\chi(Q+q,\omega)} = \frac{1}{\chi(Q)} + A q^2 - i C \frac{\omega}{q^{\theta}} ,
\label{eq:1}
\end{eqnarray}
where $\theta =1$ and $0$ for $ Q = 0$ and $Q \neq 0$, respectively. 
For non-interacting systems $A$, $C$ and $1/\chi_0 (Q)$ are calculated from the given band structure~\cite{30}
and RPA result is simply given by $1/\chi_{\rm RPA} (Q) = 1/\chi_0 (Q) - 2 U$, $U$ being the on-site interaction 
constant and the susceptibility is defined without a factor $4 \mu_{\rm B}^2$. 

In the self-consistent renormalization theory the renormalized values of $A$ and $C$ usually stay constant 
around the magnetic instability while the temperature dependence of $1/\chi(Q)$  is strikingly renormalized by 
the mode-mode coupling effects, i.e., from the Sommerfeld expansion to the quantum critical and Curie-Weiss behaviors.

Leaving derivations of the theory for ref. ~\cite{4,5}
we show here only the equations for $\chi (Q)$.
\begin{eqnarray}
     \frac{1}{\chi(Q)} &=& \frac{1}{\chi_0 (Q)} -2 U +\frac{5}{3} F_{Q} \sum_{\alpha = x,y,z} m_{\alpha}^2 ,
\nonumber \\
      m_{\alpha}^2 &=& \frac{2}{\pi} \int^\infty_0 {\rm d}\omega \left( \frac{1}{2} +\frac{1}{e^{\omega/T}-1} \right)
	                            {\sum_{q}}^\prime {\rm Im} \chi_\alpha (Q+q,\omega)
\label{eq:2}
\end{eqnarray}
where $F_Q$ is the mode-mode coupling constant for those with the wave vectors around $Q$. 
Eqs. (\ref{eq:1}) and (\ref{eq:2}) should be solved self-consistently for $1/\chi(Q)$.

For convenience we introduce here a reduced inverse susceptibility and the following parameters:
\begin{eqnarray}
     y &=& 1/ 2 T_A \chi (Q) ,
\label{eq:3}
\\
      T_A &=& A q_{\rm B}^2 /2 , ~ T_0 = (A/C) q_{\rm B}^{2+\theta} / 2 \pi,
\nonumber \\
      y_0 &=& 1/2  T_A \chi(Q,T=0) , ~ \mbox{for a paramagnetic ground state},
\nonumber \\
              && - F_Q p_Q^2 / 8  T_A , ~ \mbox{for a magnetically ordered ground state},
\nonumber \\
     y_1 &=& 5 F_Q T_0 /T_A^2
\label{eq:4}
\end{eqnarray}
where $p_Q$ is the ordered moment in $\mu_{\rm B}$ per magnetic atom, 
$q_{\rm B} =(\frac{2D \pi ^{D-1}}{v_0})^{1/D}$ is the effective Brillouin zone boundary vector, 
$D$ dimensionality and $v_0$ is the volume per magnetic atom. The dynamical susceptibility is now written as
\begin{eqnarray}
     \frac{1}{2 T_A \chi(Q+q, \omega)} = y+x^2 -i \frac{\nu}{x^\theta} ,
\nonumber 
\end{eqnarray}
with
\begin{eqnarray}
     \nu =\omega /2\pi T_0 , ~ x=q/q_{\rm B},
\label{eq:5}
\end{eqnarray}
and
\begin{eqnarray}
     y=y_0 +\frac{D}{2} y_1 \int_0^{x_c} {\rm d}x x^{D+\theta -1} \left[ \ln u -\frac{1}{2u} -\psi(u) \right],
\nonumber 
\end{eqnarray}
with
\begin{eqnarray}
     u=x^\theta(y+x^2)/t,~ t=T/T_0.
\label{eq:6}
\end{eqnarray}

We summarize the results of this theory as follows~\cite{4,5}:
\begin{enumerate}
\item[(1)] $\chi(Q)$ obeys the Curie-Weiss law with the origin different from the traditional
        local moment mechanism. 
\item[(2)] The quantum critical behaviors are obtained as shown in Table~\ref{tab1}.  The 3D and
        2D results were obtained in 1970's and 1990's, respectively. Many of the results 
        were confirmed by experimental investigations from 1970's to 1990's. Recent 
        renormalization group studies of the same problem lead to the same QC indices
        as the above results~\cite{31,32}.
\item[(3)] The Curie and N\'{e}el temperatures of 3D magnets are given by
        \begin{eqnarray}
             T_{\rm C} = 0.1052 p^{3/2} T_A^{3/4} T_0^{1/4},
        \nonumber \\
             T_{\rm N} = 0.1376 p_Q^{4/3} T_A^{2/3} T_0^{1/3},
        \label{eq:7}
        \end{eqnarray}
        These results were confirmed qualitatively(chemical and physical pressure
        dependence) and quantitatively by experimental investigations since 1970's.	
\item[(4)] Expressions for various physical quantities with the 4 SCR parameters are given
        and are applicable to a wide range of temperature including both QC and CW  regimes. 
\end{enumerate}
\begin{table}
\begin{center}
\begin{tabular}{c|cc|cc}\hline
~ & \multicolumn{2}{|c|}{\small ferro. ($Q=0$)}  &  \multicolumn{2}{|c}{\small antiferro.} \\
~ & {\small 3D} & {\small 2D} & {\small 3D} & {\small 2D}  \\ \hline
{\small $\chi_Q^{-1}$}     & {\small$T^{4/3}  \to \mbox{CW}$} 
                                         & {\small $-T \ln T \to \mbox{CW}$} 
                                         & {\small $T^{3/2} \to \mbox{CW}$} 
                                         & {\small $-T / \ln T \to \mbox{CW}$}  \\ 
{\small $C_{\rm m} /T$} & {\small $-\ln T$}                               & {\small $T^{-1/3}$}
                                        & {\small ${\rm const.} -T^{1/2}$}      & {\small $-\ln T$} \\
{\small $1/T_1$}            & {\small $T \chi$}                               & {\small $T \chi^{3/2}$}
                                        & {\small $T \chi_{Q}^{1/2}$}              & {\small $T \chi_{Q}$} \\
{\small $R$ }                   & {\small $T^{5/3}$}                            & {\small $T^{4/3}$}
                                        & {\small $T^{3/2}$}                            & {\small $T$} \\ \hline
\end{tabular}
\end{center}
\caption{Quantum critical behaviors of physical quantities.}
\label{tab1}
\end{table}
For  early investigations on transition metals and their compounds we refer to ~\cite{4}. 
Recently non-Fermi liquid properties of certain heavy $f$-electron systems are investigated intensively 
and are analyzed in terms of the QC spin fluctuation of 3D or 2D character depending on substances~\cite{33}.

\section{Anomalous normal state properties of the high-$T_{\rm c}$ cuprates}

From the beginning of investigations the anomalous or non-Fermi liquid properties in the normal 
state of the cuprates were represented by the $T$-linear electrical resistivity in a wide temperature range, 
anomalous temperature dependences of the Hall coefficient  $R_H$ and NMR $T_1$ , $R_H$ and $1/T_1T$  
showing the Curie-Weiss behaviors, and the $\omega$-linear relaxation rate of the optical conductivity , 
etc. In an early stage of investigations all of these properties except for the Hall coefficient were successfully 
analyzed in a consistent manner in terms of the above discussed spin fluctuation theory. 
The phenomenological parameters of the theory were determined from analyses of the resistivity and $T_1$. 
By using thus determined dynamical susceptibility a parameter free comparison between theory and experiment 
were performed successfully on the optical conductivity of YBa$_2$Cu$_3$O$_7$~\cite{34}. 

The Hall effect was studied recently by using the Kubo-Eliashberg formalism on the Hubbard model 
with appropriate transfer matrices and electron occupations~\cite{35}. 
The RRPA (renormalized random phase approximation)
or FLEX approximation was used for the spin fluctuations and the importance of 
the vertex corrections in dealing with strongly anisotropic scatterings was emphasized.  
The results for $R_H$ were consistent with experimental results both for the hole- and electron-doped cuprates; 
$R_H$ in the former is positive and shows a Curie-Weiss behavior  while the latter decreases with lowering 
temperature from a positive value at high temperature and changes sign, the decrement showing a Curie-Weiss behavior. 

In the second stage of investigations the pseudo-gap phenomena observed in under-hole-doped cuprates have called 
attention of investigators. This may be regarded as low temperature corrections to the non-FL behaviors as 
discussed in the first stage. We will discuss this problem later in this article and now move to the problem 
of superconductivity induced by the spin fluctuations. 

\section{Superconductivity induced by antiferromagnetic spin fluctuations}

\subsection{Theories using the parametrized spin fluctuations}

Since the anomalous normal state properties were explained by the spin fluctuation theory and the dynamical 
susceptibilities were estimated quantitatively from analyses of the experimental results, 
next problem is naturally to see if the same spin fluctuations can explain the high $T_{\rm c}$ superconductivity 
of the cuprates.

It was shown in earlier studies by using a weak coupling theory ~\cite{36,37}
and a strong coupling theory~\cite{18}
that the antiferromagnetic spin fluctuations can give rise to a $d$-wave superconductivity.  
The spin fluctuation induced superconductivity of the cuprates 
were studied first by using the weak coupling theory~\cite{10,11}
and then the strong coupling theory~\cite{12,13}
which is more appropriate. The results showed the  $d$-$x^2-y^2$ symmetry of the order parameter 
and values of $T_{\rm c}$ consistent with the observed values. 

After the unconventional $d$-wave symmetry of the order parameter was confirmed experimentally 
the relations between the spin fluctuation parameters and $T_{\rm c}$ were investigated. 
The most remarkable result is that $T_{\rm c}$ is roughly proportional to $T_0$, 
the energy spread of the spin fluctuations. Dependences of $T_{\rm c}$ on the other parameters, $T_A$, $y_0$, 
and doping concentration are relatively weak. 

The effect of dimensionality was also studied. In favorable cases the values of $T_{\rm c}$ 
can be comparable in 3D and 2D systems and are roughly proportional to $T_0$ , 
although the dependences of $T_{\rm c}$ on the other parameters are much more significant in 3D than in  2D 
systems and 2D seems to be generally more favorable than 3D for superconductivity~\cite{17,38}. 
In the same context FLEX approximation was applied to a nearly half-filled Hubbard model with a simple cubic 
lattice and the calculated $T_{\rm c}$ was found to be extremely small~\cite{39}.  
A more recent study of a spatially anisotropic Hubbard model with a transfer parameter interpolating between 
2D and 3D systems show that $T_{\rm c}$ varies only slowly on the 2D side of the parameter values and then 
decreases rapidly as one approaches the cubic limit~\cite{40}.

\begin{figure}[!ht]
\begin{center}
\includegraphics[width=0.6\textwidth]{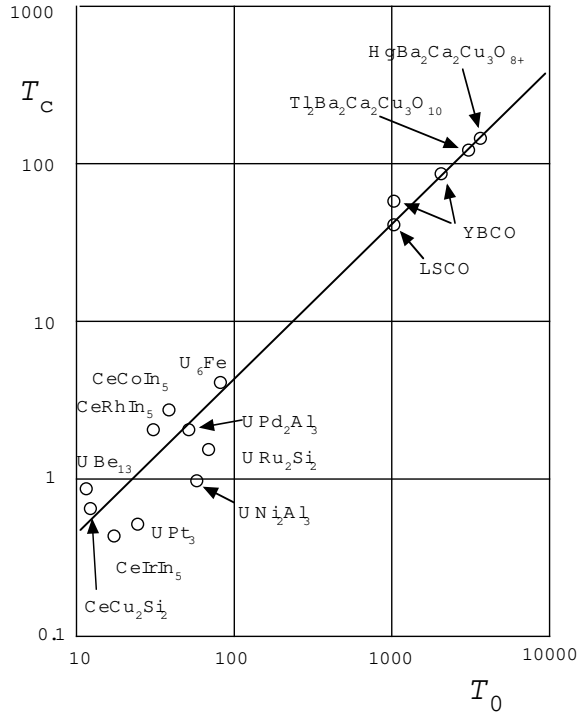}
\end{center}
\caption{Transition temperature of unconventional superconductors plotted against $T_0$, 
             the characteristic temperature indicating the energy spread of spin fluctuation.}
\label{fig2}
\end{figure}

We show in Fig.~\ref{fig2} the plots of experimental values of $T_{\rm c} $
against $T_0$ for cuprates and heavy electron superconductors~\cite{17}. 
The plots come around a straight line, suggesting the common origin of superconductivity in these groups of systems.

\subsection{Theories based on  the microscopic models}

Fully microscopic calculations of spin fluctuation-induced superconductivity were carried out on 
the Hubbard and the $d$-$p$ models by using the FLEX approximation~\cite{18,19,20,21,22,24,25}, 
the 3rd order perturbation theory~\cite{23}, 
and variational Monte Carlo method~\cite{41}. 
The results may be summarized as follows~\cite{5}:
\begin{enumerate}
\item[(1)] Superconductivity of $d$-$x^2-y^2$ symmetry is obtained in a fairly large range of
        intermediate values for $U/t$ . The calculated values of $T_{\rm c}$ are of reasonable 
        magnitude compared with experiment.
\item[(2)] Similar results are obtained both for the Hubbard and $d$-$p$ models.
\item[(3)] For the superconducting state, the neutron resonance peak observed in YBCO is
        explained by the FLEX calculations and by RPA calculations assuming the BCS
        ground state. Also a peak-dip-hump structure in the one electron spectral density
        as observed by angle-resolved photoemission experiment is explained.
\item[(4)] Observed differences between the phase diagrams for the electron-doped and
        hole-doped cuprates are explained at least qualitatively; a substantially larger 
        concentration range of AF phase and smaller values of $T_{\rm c}$ in the former as
        compared with the latter~\cite{42,43}.  
        The calculated ratio of the SC gap and $k_{\rm B} T_{\rm c}$
        for the former is about half of that in the latter in accord with measured results. 
        This difference may be related with the van Hove singular points which are close to
        the hot spots, where scatterings due to AF spin fluctuations is particularly strong, in 
        the hole-doped systems but not in the electron-doped systems. 
\end{enumerate}

It may be worth while to remark that the FLEX approximation is the simplest possible 
approach for self-consistently renormalized spin fluctuations and the above success strongly 
favors the spin fluctuation mechanism for the superconductivity in the cuprates. 
It is also remarkable that the Hubbard model in the intermediate coupling regime seems 
to be a good model for the cuprates if one chooses the transfer matrices so as to reproduce the observed 
Fermi surface for each substance.

\section{2D-organic superconductors and the high-$T_{\rm c}$ cuprates \\
    		  {\it An extended phase diagram for the Hubbard model }}

An organic system  $\kappa$-(BEDT-TTF)$_2$CuN(CN)$_2$Cl is an antiferromagnetic insulator at ambient pressure.
Under applied pressure it undergoes a weak first order transition, at $p = 200 {\rm bar}$, 
into a metallic state which shows superconductivity below $T_{\rm c} =13{\rm K}$.  
To a good approximation this substance is considered to be described by a half-filled Hubbard model consisting 
of antibonding dimer orbitals arranged in a square lattice. As for the transfer integrals we may take $-t$ 
between the nearest neighbors and $t^\prime$ between the second neighbors in one of the diagonal directions. 

This model was studied with the FLEX approximation and the results showed the superconducting order parameter of 
$d$-$x^2-y^2$ symmetry and reasonable values for $T_{\rm c}$~\cite{26,27,28,44}. 
Since the approach here is just the same as the one discussed in the preceding section for the cuprates, 
the common success naturally indicates the same origin of superconductivity in these systems.

The above results are of particular interest since the organic compound is found in the metallic side 
of the Mott transition while the high-$T_{\rm c}$ cuprates are doped Mott insulators. 
The former is clearly in the intermediate coupling regime and existing approaches from the strong coupling 
limit does not seem to work.

\begin{figure}[!ht]
\begin{center}
\includegraphics[width=0.6\textwidth]{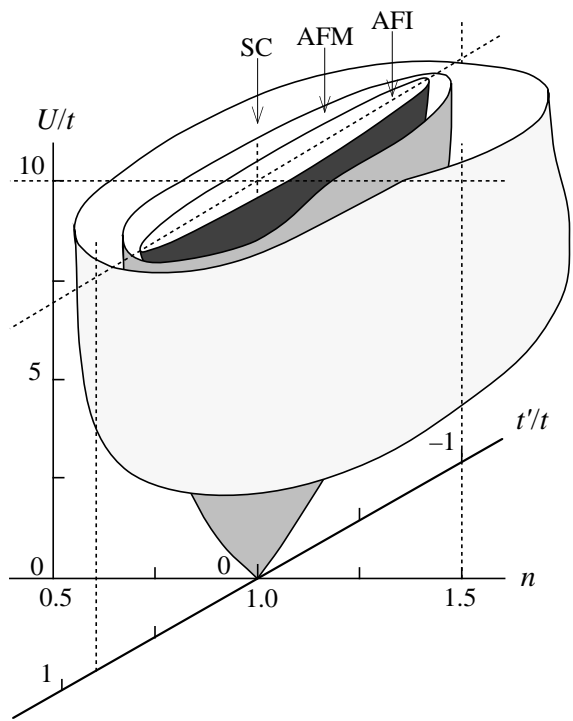}
\end{center}
\caption{The phase diagram of a nearly half-filled Hubbard model in a three dimensional parameter space: 
              $U/t$-$t^\prime/t$-$ n$. Superconducting (SC) and antiferromagnetic (AFM) instability surfaces are indicated. 
			  The boundary between antiferromagnetic metal(AFM) and insulator(AFI) should depend on the impurity potential 
			  and is sketched here just to show general idea of its location.}
\label{fig3}
\end{figure}

In order to see the situations more clearly a phase diagram was calculated in a parameter space 
of the Hubbard model taking the following quantities for the 3 axes of the parameter space: $U/t$, $t'/t$ and $n-1$, 
$n$ being the number of electrons per site~\cite{29}. 
The FLEX approximation was employed and the results were extrapolated to $T = 0$ with the use of Pad\'{e} approximants. 
The calculated phase diagram is shown in Fig. ~\ref{fig3}. We see that the superconducting states of cuprates and 
those of the organic compounds are found in the same connected region of  a superconducting phase in this phase diagram. 
%

\section{Magnetic instability vs. Mott transition  \\
    		  {\it Which is more important for high-$T_{\rm c}$ ? }}

Among the mechanisms of superconductivity in strongly correlated electron systems the one for 
the heavy electron systems is considered mainly due to the spin fluctuations, 
although additional contributions of orbital and charge fluctuations are under investigation~\cite{45}. 
For the 2D organic compounds the only mechanism explicitly worked out so far seems 
to be the spin fluctuation mechanism.

For the high-$T_{\rm c}$ cuprates, on the other hand, many different mechanisms have been proposed. 
In addition to the approaches as discussed in the above, various approaches from the strong coupling limit 
based on the $t$-$J$ model have been pursued extensively~\cite{46,47}  
Since the $t$-$J$ model is an approximation to the Hubbard model from the strong coupling limit it may be 
pertinent here to discuss on the phase diagrams of the Hubbard model and see where in the parameter space the 
high-$T_{\rm c}$ cuprates are really located.

\begin{figure}[!ht]
\begin{center}
\includegraphics[width=1.\textwidth]{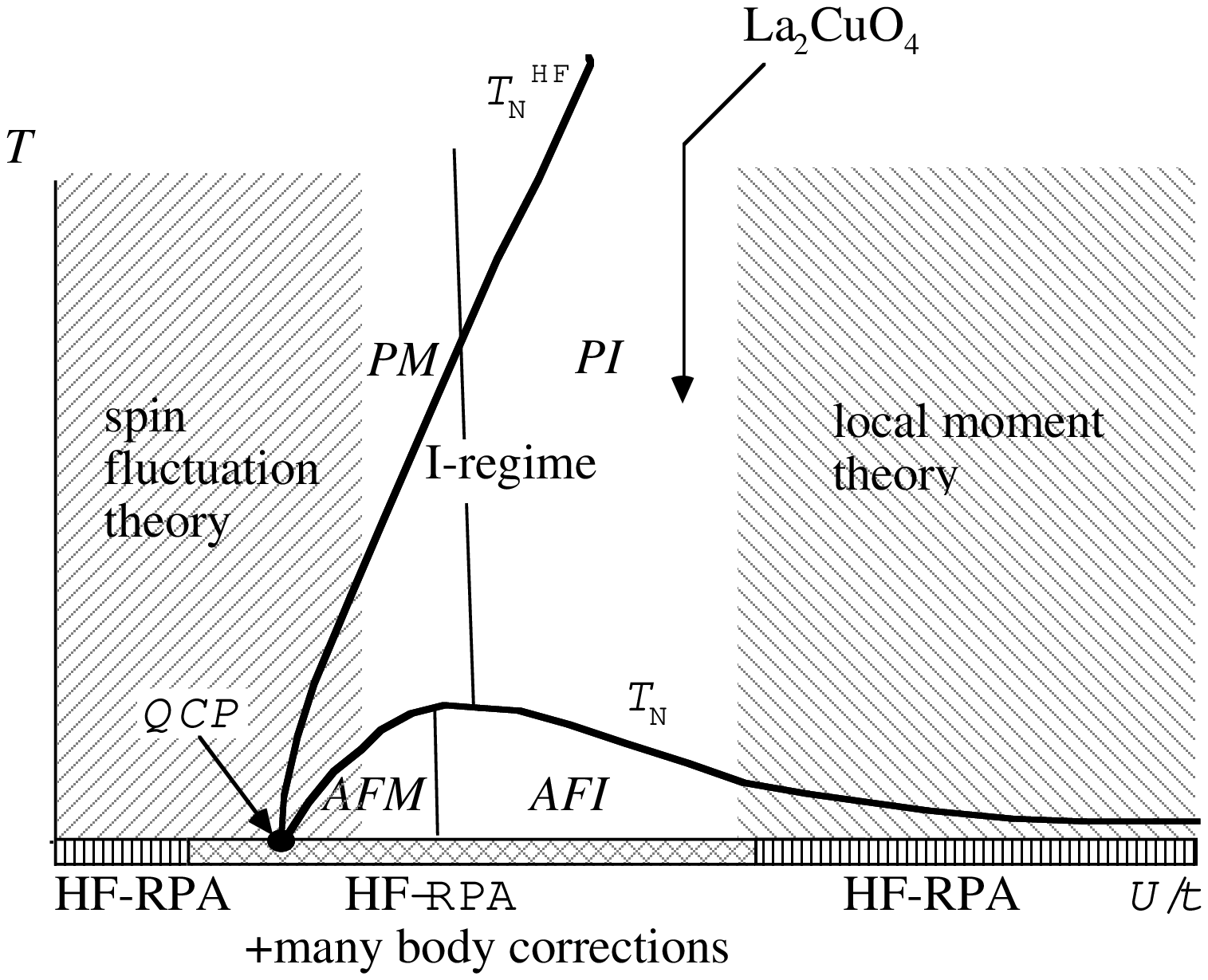}
\end{center}
\caption{A sketch of the phase diagram for the half-filled Hubbard model. 
              Appropriate theoretical approaches are indicated in the corresponding sections of the phase diagram.}
\label{fig4}
\end{figure}

Let us first look at a phase diagram of a half-filled Hubbard model including the magnetic QC point, 
an antiferromagnetic phase and a metal to insulator transition, Fig. ~\ref{fig4}. 
Appropriate ways of approaches for different sections of the phase diagram are also indicated in the figure. 

We first emphasize that in the ground state the Hartree-Fock(HF)-RPA is correct in the strong coupling regime where 
the ordered moment is enough polarized so that the correlation effect is insignificant; 
two electrons with opposite spins seldom encounter with each other. The spin wave dispersion calculated by RPA 
in the antiferromagnetic ground state agrees, in the limit of small $t/U$, precisely with the one calculated from
the Heisenberg model with the Anderson kinetic superexchange interactions~\cite{4}. 
Since the HF-RPA is known to be correct in the weak coupling limit it should make 
a fair interpolation between the strong and weak coupling limits in the ground state. 
In the intermediate coupling regime, where the electron-electron correlations are most significant, 
it is a reasonable approach to start from HF-RPA and to make many body corrections. 
The self-consistent theory of renormalized spin fluctuations may be an example of possible approaches.

At finite temperatures, however, HF-RPA is not at all a good approximation in any regimes of present interest. 
For the Mott insulator phase with large $U/t$ the local moment model with the Anderson kinetic 
superexchange ($t/U$ expansion) is valid while for weak and intermediate $U/t$ just covering 
the antiferromagnetic instability the SCR spin fluctuation theory is considered to be an adequate approach. 
For the intermediate regime between these two including the Mott transition no really satisfactory approach 
at finite temperatures is known so far. 

We note that 2D organic superconductors are on the metallic side of the first order Mott transition and 
is considered to be located around the hidden antiferromagnetic instability and thus are expected to be well 
treated by the spin fluctuation theory.

Next we discuss doped Mott insulators. If we neglect the impurity potential the ground state of 
a doped Mott (antiferromagnetic) insulator  is an antiferromagnetic metal. 
On doping we first have small Fermi surfaces which expand with increasing doping concentration 
at the expense of reduced gap area. Then we have a large Fermi surface with small gap area and finally have a 
paramagnetric metal through the QC point where the AF gap just vanishes. 

\begin{figure}[!ht]
\begin{center}
\includegraphics[width=1.\textwidth]{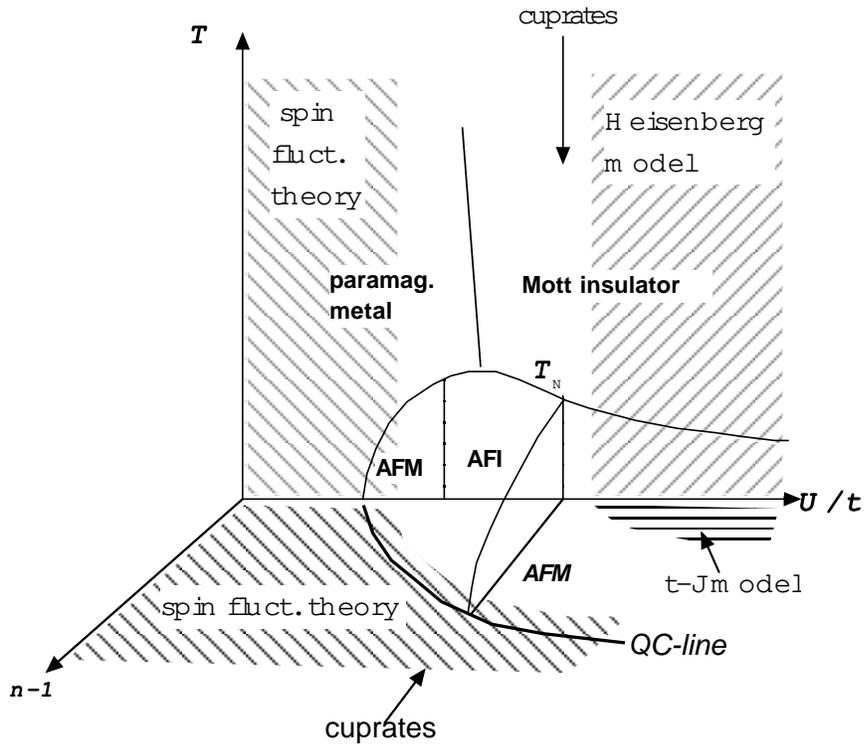}
\vspace*{-2.cm}
\end{center}
\caption{A sketch of the phase diagram for the nearly half-filled Hubbard model. 
              Appropriate approaches are indicated in the corresponding sections.}
\label{fig5}
\end{figure}

As is seen in Fig. ~\ref{fig5} the magnetic instability or QC point should make a continuous line in a $U/t$  
against n plane extending from the half-filled ($n = 1$) case with relatively small $U/t$ toward larger $U/t $ 
and $n \neq 1$. The spin fluctuation theory is an approach along the QC line from the side of weaker coupling and 
is expected to cover at least the intermediate coupling regime. As a matter of fact the values for $U/t$ 
used in the above calculations are around $4 \sim 8$ and thus $U$ is about $0.5 \sim 1$ times the band width. 
This is consistent with the value estimated from photoemission experiments~\cite{48}. 

According to recent analyses of the spin wave energy dispersion in antiferromagnetic La$_2$CuO$_4$ , 
a parent of high-$T_{\rm c}$ cuprates, as measured by neutron scattering studies~\cite{49}
the dispersion in the entire Brillouin zone is well reproduced by an RPA calculation 
on the Hubbard model with $U/t = 6$~\cite{50}.  
In order to get a reasonable fit with the local moment model one needs to consider the higher order terms 
in the $t/U$ expansion including the ring exchange terms~\cite{49}. 
This fact seems to evidence the intermediate coupling nature of the cuprates and supply another support for 
the applicability of the spin fluctuation theory to the high-$T_{\rm c}$ problems and explain the reason for its success. 

On the other hand the approaches based on the $t$-$J$ model with a few terms of Anderson superexchange interactions 
might not even cover the parent insulator state. 
Applicability of this approach as an approximation to the Hubbard model may also be limited to a very low doping 
concentration range. For larger doping concentrations it may take us into a different world from the one described 
by the Hubbard model.

\section{Pseudo-gap phenomena}

Pseudogap phenomena were first observed in NMR $T_1$ measurements on underdoped YBCO[51, 52] 
as a pseudo-gap in spin excitations. The measurements were then extended to transport properties, 
one electron spectral density as observed by ARPES and  tunnelling experiments and various theoretical mechanisms 
were proposed [53]. 

According to the FLEX calculations $T_{\rm c}$ generally tends to increase with lowering doping concentration until it 
reaches the AF phase boundary calculated introducing  weak 3-dimensionality. This behavior is consistent with the 
experimental results on the electron-doped cuprates. For the hole-under-doped cuprates, however, 
experimental results seem to suggest that there is some additional mechanism to suppress both the AF and SC orderings, 
giving rise to the pseudo-gap at the same time. 

Among various possibilities proposed so far one natural direction from the present point of view may be to consider 
superconducting fluctuations($p$-$p$ channel) in addition to the AF spin fluctuations($p$-$h$ channel). 
The SC fluctuation is expected to be particularly significant in 2D systems with short coherence lengths. 
Here we refer to an investigation reported recently[54-56]. 
The SC fluctuations are taken into account within the $t$-matrix approximation together
with the FLEX spin fluctuations and the equations are solved self-consistently 
with still further approximation  of expanding the $t$-matrix in $q^2$ and $\omega$ near the SC critical point. 
The results include many attractive features: 
(1) In the under-hole-doped regime $T_{\rm c}$ decreases from the FLEX value and the value of $T_{\rm c}$ decreases 
with lowering doping concentration. The AF phase tends to be suppressed at the same time. 
(2) Pseudogap behaviors in the one electron spectral density and reduction of the NMR relaxation rate $1/T_1T$ 
are obtained. 
(3) Observed pseudo-gap behaviors in the transport properties: electrical resistivity, Hall coefficient, 
magneto-resistance, thermo-electric power, and the Nernst coefficient are all explained consistently. 
(4) Calculated effects are significant in the hole-underdoped cuprates but not in the electron-doped 
cuprates in accord with experiment. 

In order to make this scenario convincing it is desired to extend the theory to still lower 
doping concentrations and clarify the phase diagram
near the AF insulator phase.

\section{Conclusion}

We have summarized the present status of the theories of the spin fluctuation mechanism in explaining anomalous or 
non-Fermi liquid behaviors and unconventional superconductivity in strongly correlated electron systems around 
the magnetic quantum critical point, referring to the high-$T_{\rm c}$ cuprates, 
2D organic compounds and heavy electron systems.  So far the spin fluctuation theories  
seem to be successful in explaining at least essential parts of the problems, 
indicating in particular that the spin fluctuation is the common origin of superconductivity in these systems. 

As for the high-$T_{\rm c}$ cuprates at least the central part of the SC phase seems to be approached by 
the spin fluctuation theory. It was fortunate that the high-$T_{\rm c}$ cuprates are in the intermediate 
coupling regime and are close to the magnetic instabilities. 
It still remains to describe all the hole-underdoped regime lying between 
the AF insulator phase and the optimal concentration regime, where the pseudo-gap phenomena are still 
controversial indicating the need for another mechanism
in addition to the antiferromagnetic spin fluctuations.

\section*{Acknowledgements}

The author wishes to thank K. Ueda for critical reading of the manuscript 
and H. Kondo for discussion and help in preparing the manuscript.
	 

\end{document}